\documentclass[11pt, a4paper]{article}
\usepackage{multirow}
\usepackage[centertags]{amsmath}
\usepackage{amsfonts}
\usepackage{amssymb}
\usepackage[dvips]{graphicx}
\usepackage{multicol}
\usepackage{amsthm}
\usepackage{pdfpages}
\usepackage{newlfont}
\usepackage{textcomp}
\usepackage{geometry}
\usepackage{subfigure}
\usepackage{cite}
\usepackage{authblk}
\usepackage{xcolor}
\newlength{\defbaselineskip}
\newcommand{\setlinespacing}[1]%
           {\setlength{\baselineskip}{#1 \defbaselineskip}}

\theoremstyle{plain}

\theoremstyle{definition}

\theoremstyle{remark}

\numberwithin{equation}{section}

\title{Effects of Competition and Cooperation Interaction between Agents on 
Networks in Presence of a ``Market Capacity".}
\author[1]{A. Sonubi }
\author[1]{A. Arcagni }
\author[1]{S. Stefani }
\author[2,3,4] {M. Ausloos }
\affil[1]{ Scuola di Economia e Statistica,  Department of Statistics and 
Quantitative Methods, Universit\`a degli Studi di  Milano-Bicocca, I-20126 
Milano,  Italy.}
\affil[2]{ School of Management, University of Leicester, %University Road, 
Leicester  LE1 7RH, UK \\$e$-$mail$ $address$: ma683@le.ac.uk
}  
\affil[3]{ eHumanities
group\footnote{Associate Researcher}, Royal Netherlands
Academy of Arts and Sciences (NKVA),  Joan Muyskenweg 25, 1096 CJ
Amsterdam, The Netherlands %\\ $e$-$mail$ 
$address$:marcel.ausloos@ehumanities.knaw.nl 
}
\affil[4]{  GRAPES\footnote{Group
of Researchers for Applications of Physics in Economy and Sociology}, 
  rue de la Belle Jardiniere 483, B-4031, Angleur, Belgium
    \\$e$-$mail$ $address$: marcel.ausloos@ulg.ac.be
}

\begin{document}
\maketitle
\noindent \begin{abstract}\noindent A network effect is introduced taking into 
account competition, cooperation and mixed-type interaction amongst agents along 
a generalized Verhulst-Lotka-Volterra model. It is also argued that the presence 
of a ``market capacity" enforces an indubious limit on the agent's size growth. 
The state stability of triadic agents, i.e., the most basic network plaquette, 
is investigated analytically for possible scenarios, through a fixed point 
analysis. It is discovered that: (i) ``market" demand is only satisfied for full 
competition when one agent monopolizes the market; (ii) growth of agent size is 
encouraged in full cooperation; (iii) collaboration amongst agents to compete  
against one single agent may result in the disappearance of this single agent 
out of the market, and (iv) cooperating with two ``rivals" may become a growth 
strategy for an intelligent agent.
%\noindent \textbf{Keywords}: \emph{Generalized Verhulst-Lotka-Volterra Model, 
%Network, Triads, Market Capacity.}
\end{abstract}

%\subsubsection*{HIGHLIGHTS}

%\begin{enumerate}
%  \item  An extension of  a  generalized Verhulst-Lotka-Volterra model \cite{comp} is proposed, introducing a network effect taking into account competition, cooperation or a mixed-type interaction amongst agents respectively.
  %\item The stability of triadic agents is investigated analytically for all possible scenarios.
  %\item Through simulations it is shown that:
 % \begin{itemize}
   % \item
   %\item In  the fully cooperative scenario, the growth of the market share of all the agents is encouraged.
   % \item  , after some time. However, this elimination  is not possible if the intensity of interaction is very low (i.e. the single agent is ``extremely bigger" in size than the collaborating agents)
    %\item Cooperating with two ``rivals" may become a growth strategy for an agent.
 %\end{itemize} \end{enumerate}
\section{INTRODUCTION}
Complex Networks and Multi-Agent Systems, are very active fields of research. 
They entail the study of networks of interacting agents whose structure is 
irregular, complex and dynamically evolving with time \cite{linked, Barabasi, 
cn, kwapiendrozdzPR12}. Most real systems in biology, chemistry, engineering, 
socioeconomics, are made up of millions and millions of interacting agents (i.e 
atoms, electrons, people as the case may be) which account for fluctuations in 
the overall behaviour of the system. Over the years, it has also been discovered 
that non-linear mathematical models  replicate the dynamics of real systems 
better than the linear ones \cite{djvu, strogatz}. An example of such former 
models is the Lotka-Volterra model  \cite{Lot,vot},  and its generalisation 
\cite{bank,ecology,ecom,low, Ma,Ada,PhA389.10NKVZIDMAideol,ACSmigration,coop, 
comp, rep}, which is also known as the prey-predator model \cite{pekalskirevPP}.

The Lotka-Volterra model has been used in various ways to model complex 
systems\cite{bank,ecology,ecom,low, Ma, 
Ada,PhA389.10NKVZIDMAideol,ACSmigration}. For instance, in \cite{ecom}, the 
model was used in e-commerce web sites in a competitive scenario to explore its 
effects and the characteristic of ``rich gets richer'' on Internet economy, and 
the ``winner-takes-all" phenomenon. This was an improvement on \cite{ Ma, Ada} 
where interacting agents where discovered to be ``sharing the market". The 
Verhulst-Lotka-Volterra model was  generalized in \cite{comp} through the 
introduction of a non-linear, symmetric interaction function, used to 
investigate a competitive scenario. This resulted in a self-organising 
clustering of agents, which were either chaotic or/and non chaotic in nature as 
a result of their dependence on the agent's size and  their initial state 
conditions. Cooperative scenarios examined in \cite{coop} were modeled 
individually without imposing any ``market condition". This resulted in the 
growth of different clusters of interacting agents beyond their capacity   in 
contrast to the competitive case. A complex network representation of the 
competitive and the cooperative scenarios was presented in \cite{rep}, in order 
to give another description of some generalized Verhulst-Lotka-Volterra model.

This present paper offers new insights on the effects of competitive and/or 
cooperative interaction in a multi-agent system when $n$ agents have some common 
resources to share in the ``market".

The main contribution of our paper is two folds: firstly, we generalize the 
Verhulst-Lotka-Volterra model by introducing a network effect through an 
undirected but weighted graph. The weights suitably represent competition or 
cooperation. The elements of the resulting adjacency matrix replaces the 
strength parameter in the interaction function of the\cite{comp, coop} model in 
order to enable a mixed-type of interactions, i.e., having a system in which 
competitive and cooperative scenarios are considered to occur simultaneously 
amongst the various  interacting agents.

Secondly, we introduce a market capacity in the model to replace individual 
agent's capacity:  this is  a realistic constraint, i.e., the maximum level 
which all the agents may reach in the market, like in Verhulst model of limited 
population growth \cite{v1,v2}- thereby enforcing a natural (endogenous) limit 
on agent's size growth.

Such competition, cooperation and  mixed-type of interactions are analyzed below 
for triad interacting agents through the evaluation of the eigenvalues of  the 
relevant Jacobian matrix computed at corresponding fixed points, in order to 
investigate the system stability.  This triad system has been chosen as the most 
simple yet complex enough as  representative of basic networks 
\cite{EPJB86.13GRMAcomplex}.   Notice that the model goes well away for the 2-player prisoner (dilemma) game. 

Thus, a mixed-type of interactions between agents, made possible through the 
network effect in the generalized Verhulst-Lotka-Volterra model, markedly 
differs  from previous works \cite{comp,rep,coop}, and seems more realistic.

This paper is organized as follows: in Section 2, the generalized Verhulst-Lotka-Volterra model is discussed. Section 3 contains the outline of the mathematical model used in this paper. The fixed point analysis and state stability are investigated in Section 4. In  Section 5,  initial size conditions and convergence to a steady state of the triad interacting agents are further illustrated through simulations. We demonstrate the presence of growth and/or 
decay effects in various scenarios, - sometimes rather complex, but  {\it a posteriori} understandable. The paper is concluded in Section 6. It appears that the model is suitable for describing not only  an ``economic market" but other 
agent based cases such as co-autorship, or more generally team working, or any other small or large network   based  complex systems.

\section{THE GENERALIZED VERHULST-LOTKA-\\VOLTERRA MODEL}
\noindent A generalized Verhulst-Lotka-Volterra Model introduced in \cite{comp} 
is given by:
\begin{equation}
 \dot{s_{i}} = \alpha_{i} s_{i}(\beta_{i} - s_{i}) - \sum_{i \neq j} 
\gamma(s_{i},s_{j}) s_{i} s_{j}, \quad \quad \quad i=1,\ldots,n, \label{2a}
\end{equation}
where $s_{i}$ is the size of agent $i$ such that $0\leq s_{i} \leq 1$; 
$\dot{s_{i}}$ is its time derivative; $\alpha_{i}$ is agent's growth rate if no interaction is present, $\beta_{i}$ is the agent's maximum capacity and 
$\gamma(s_{i},s_{j})$ is the interaction function. The first term is a Verhulst-like term \cite{v1,v2} and the others stem from the Lotka-Volterra model 
\cite{Lot,vot}.

The interaction function $\gamma(s_{i},s_{j})$  is defined by
 \begin{equation}
  \gamma(s_{i},s_{j}) = K \exp \left[- \left(\frac{s_{i}-
s_{j}}{\sigma}\right)^{2}\right]; \label{2b}
\end{equation} it is a continuously differentiable function that allows a proper theoretical analysis of the system dynamics leading to conclusions which will appear to be likely model independent. The positive parameter $\sigma$ controls 
or scales the intensity of agent size similarity and the parameter $K$ determines the scenarios of agent's interaction.

The model is used in \cite{comp} to analyze a system of $n$ agents in competition for some common resources with competition becoming more aggressive 
between agents with similar size. This is because as $|s_{i} - s_{j}|\rightarrow 0$, the interaction function $\gamma(s_{i}, s_{j}) \rightarrow K$ for a constant parameter $\sigma$, its maximum value. The  competition weakens  when  agents  
have distinctly different sizes, thereby suggesting  a peer-to-peer competition modelling.

For the peer-to-peer interaction system,  presented in \cite{comp,coop}, a strength parameter $K$  was  introduced: $K>0$ was considered to show  the presence of competition in the market, while $K<0$ implied cooperation. Under 
the cooperative scenario, the interaction function is  defined  in the interval $-K < \gamma(s_{i},s_{j}) < 0$. In order to avoid complexity and instability of 
the system, the value of $K$ was chosen carefully through fixed point analysis of agents with equal sizes\cite{coop}, whose eigenvalues are
\begin{equation}
 \lambda_{1, \ldots, n} = \frac{K-1}{1 +(n-1)K}, \label{2c}
\end{equation}
 so that, the $K$ range interval $- \frac{1}{n-1} < K < 0$ ensures the stability of the system.
 
\section{THE MODEL}
\noindent This paper is a study of the Verhulst-Lotka-Volterra model when competition and cooperation can occur between linked agents. The market capacity $\beta$ becomes the amount of product/service sales that could be reached within a certain period of time by any agents in the ``market" \cite{capacity}. The relationship between the individual agent's maximum capacity $\beta_{i}$ and $\beta$ is given by:
\begin{equation} 
\beta_{i} = \beta - \sum_{i \neq j} s_{j}, \label{2d}
\end{equation}
\noindent so that the  initial Verhulst-Lokta-Volterra model becomes:
\begin{equation} 
 \alpha_{i} s_{i}(\beta_{i} - s_{i}) = \alpha_{i} s_{i}\left(\beta - \sum_{i= 1} 
s_{i}\right).\label{2e}
\end{equation}
In addition, interaction among agents is introduced and modeled by a matrix 
$\textbf{K}$ with elements $k_{ij} = k_{ji}$, which are zero on the diagonal, and can be +1 or -1 off the diagonal. Thus, the interaction function becomes:
\begin{equation} \gamma(s_{i},s_{j}) = k_{ij} \exp \left[- \left(\frac{s_{i}-
s_{j}}{\sigma}\right)^{2}\right]
\end{equation}
with $0< \gamma(s_{i},s_{j}) < |k_{ij}|$.

This matrix $\textbf{K}$ is the adjacency matrix of a network represented by a weighted and undirected graph. The weights are $0,-1$  and $+1$ indicate no interaction, cooperation and competition respectively.
Furthermore, we assume that there is no loop, that is, an agent cannot compete or cooperate with herself.
For some special matrices $\textbf{K}$, we obtain the model in \cite{coop,comp}. 
For instance, when
\begin{equation} \scriptstyle \nonumber \textbf{K}  \equiv \left[
                    \begin{array}{ccccc}
                     0 & 1& . & . & 1  \\
                       1 & 0&  & .&1  \\
                       .& & . &  & . \\
                       .& &  & . & .\\
                     1 & 1  & . & . & 0
                    \end{array} \right]
\end{equation}
\noindent we obtain the full competitive scenario as in \cite{comp}. For
\begin{equation} \scriptstyle \nonumber \textbf{K} \equiv \left[
                    \begin{array}{ccccc}
                     0 & -1& . & . & -1  \\
                       -1 & 0&  & .&-1  \\
                       .& & . &  & . \\
                       .& &  & . & .\\
                     -1 & -1  & . & . & 0
                    \end{array}\right],
\end{equation}
\noindent this is the full cooperative scenario \cite{coop}. In addition the 
network matrix %$K$
\begin{equation} \scriptstyle \nonumber \textbf{K}  \equiv \left[
                    \begin{array}{ccccc}
                     0 & k_{12}& . & . & k_{1n}  \\
                       k_{21} & 0&  & .&k_{2n}  \\
                       .& & . &  & . \\
                       .& &  & . & .\\
                     k_{n1} & k_{n2}  & . & . & 0\\
                    \end{array}
                  \right]
                   \end{equation}
ensures a mixed-type of interaction amongst the agents in the model, when $k_{ij} (= k_{ji})$ takes the value +1 or -1. That is, when competition and cooperation occur simultaneously amongst interacting agents in the system. For 
example, for $n=3$, in our model, we can have two agents collaborating in order to compete effectively with the third agent. This can be compared with the case of two small companies collaborating to compete against a big company which had previously monopolized the market or when two political parties merge together for the purpose of winning  over a ruling (or  over a potentially ruling) party in an election or when scientific teams or authors cooperate on some research 
topics.

Thus suppose that there exist  $n$ agents sharing some common resources. Let us assume that agents increase in size if they acquire some portion of the resources or have  their size reduced if any market portion is lost. Our 
mathematical model is defined by an $n$-dimensional differential equation:
\begin{equation}  \dot{s}_{i} = \alpha_{i} s_{i}\left(\beta - \sum_{i= 1}^{n} 
s_{i}\right) - \sum_{i \neq j} k_{ij} \exp \left[- \left(\frac{s_{i}-
s_{j}}{\sigma}\right)^{2}\right] s_{i}s_{j}, \quad \quad \quad \quad i = 1, 
\ldots ,n
\label{2f}
\end{equation}
where $s_{i}$ is the agent  size such that $0 < s_{i} \leq 1$; $\dot{s}$ is its time derivative; $\alpha_{i}$ is the growth rate of agent $i$ if no interaction is present; $\beta$ is the market capacity and $k_{ij}$ is the element of the network matrix $\textbf{K}$ which determines the interaction between agent $i$ and $j$. The interaction function $\gamma(s_{i},s_{j})$ is composed of dynamic parameters 
that result from the difference between agents in relation; the parameter $\sigma$ is a positive parameter that regulates the difference in the agent's size. Note that while $\textbf{K}$ indicates what interaction is present,  
$\sigma$ determines the ``range" of interaction of the agents.
Indeed, in contrast to the ``large"  interaction between equal size agents, the intensity of interactions between ``agents with bigger market share" and ``those with small market share"   occurs to be weak, since as $\mid s_{i} - s_{j}\mid \rightarrow \infty$, the interaction function $\gamma(s_{i},s_{j}) \rightarrow 0$; on the contrary, 
indeed, as $\mid s_{i} - s_{j}\mid \rightarrow 0$,
$\gamma(s_{i},s_{j}) \rightarrow \pm 1$ depending on $k_{ij}$,
which signifies a strong interaction between agents with similar sizes.

For the sake of simplicity, without much losing generality, it can be  assumed 
that the agents have the same dynamic properties $\alpha_{i} = 1$, while  the 
market capacity  can be $\beta = 1$. Therefore, equation (\ref{2f}) becomes:
\begin{equation}  \dot{s}_{i} =  s_{i}\left(1 - \sum_{i= 1}^{n} s_{i}\right) - 
\sum_{i \neq j}  k_{ij} \exp \left[- \left(\frac{s_{i}-
s_{j}}{\sigma}\right)^{2}\right]
s_{i}s_{j}, \quad \quad \quad \quad i = 1, \ldots ,n
\label{2g}
\end{equation}
where, as stated earlier, $k_{ij}$, $i,j=1,\ldots,n$ are elements of the interaction matrix $\textbf{K}$.
\section{FIXED POINT ANALYSIS AND STABILITY}
In this section, the fixed point analysis of the model is done in order to investigate the stability of the system for all  scenarios. A triad system of agent (i.e. $n = 3$) was chosen as a simple yet complex representative of a 
basic network, in order to illustrate some of the properties of the model analytically.

Suppose $A_{1}, A_{2}$ and $A_{3}$ are triad interacting agents with market sizes ${s}_{1}, {s}_{2}$ and ${s}_{3}$ respectively, from  equations (3.3) the system of triads becomes:
\begin{eqnarray}  \dot{s}_{1} &=&  s_{1}(1 - s_{1} - s_{2} - s_{3}) -  k_{12} 
\exp^ {- \left(\frac{s_{1}-s_{2}}{\sigma}\right)^{2}}s_{1}s_{2} - k_{13} \exp^{- 
\left(\frac{s_{1}-s_{3}}{\sigma}\right)^{2}}s_{1}s_{3},
\label{4a}\\  \dot{s}_{2} &=&  s_{2}(1 - s_{1} - s_{2} - s_{3}) -  k_{12} \exp^ 
{- \left(\frac{s_{2}-s_{1}}{\sigma}\right)^{2}}s_{2}s_{1} - k_{23} \exp^{- 
\left(\frac{s_{2}-s_{3}}{\sigma}\right)^{2}}s_{2}s_{3}, 
\label{4b}\\  \dot{s}_{3} &=&  s_{3}(1 - s_{1} - s_{2} - s_{3}) -  k_{13} \exp^ 
{- \left(\frac{s_{3}-s_{1}}{\sigma}\right)^{2}}s_{3}s_{1} - k_{23} \exp^{- 
\left(\frac{s_{3}-s_{2}}{\sigma}\right)^{2}}s_{3}s_{2}.
\label{4c}
\end{eqnarray}

The possible $\textbf{K}$-matrices for describing the  different scenarios of interaction amongst the agents $A_{1}, A_{2}$ and $A_{3}$ are:
\begin{equation} \scriptstyle \nonumber  \textbf{K}_{1} = \left[
                    \begin{array}{ccccc}
                     0& 1& 1 \\
                      1 & 0& 1  \\
                       1&1 & 0\\
                    \end{array}
                  \right], \textbf{K}_{2} = \left[
                    \begin{array}{ccccc}
                     0& 1& 1 \\
                      1 & 0& -1  \\
                       1&-1 & 0\\
                    \end{array}
                  \right],\textbf{K}_{3} = \left[
                    \begin{array}{ccccc}
                     0 & -1& -1 \\
                      -1& 0& 1  \\
                       -1&1 & 0\\
                    \end{array}
                  \right], \textbf{K}_{4} = \left[
                    \begin{array}{ccccc}
                     0 & -1& -1 \\
                      -1 & 0& -1  \\
                       -1&-1 & 0\\
                    \end{array}
                  \right]
                  \label{4d}
\end{equation}
where $\textbf{K}_{1}$ represents the matrix for a full competitive system with three interacting agents; $\textbf{K}_{2}$ is a matrix for  mixed-type of interaction system, where one agent compete with two other agents in cooperation. In this case, agent $A_{1}$ competes with agent $A_{2}$ and agent 
$A_{3}$ which are in cooperation. $\textbf{K}_{3}$ is a matrix for mixed-type of interaction system, where one agent cooperates with two other agents in competition, in this case, agent $A_{1}$ cooperates with agent $A_{2}$ and agent 
$A_{3}$ which are in competition with each other.  $\textbf{K}_{4}$ represents the matrix for a full cooperative system amongst the three interacting agents. It can be observed that other cases of the mixed-type of interaction are 
isomorphic to the ones here presented.

The fixed point analysis of the system entails the evaluation of the eigenvalues of the relevant Jacobian matrix computed at each corresponding fixed point, thus 
used to determine the stability of the system. When the real part of all the eigenvalues is negative, the system is said to be stable. If at least one eigenvalue has a positive real part, the system is unstable.
\subsection{Fully Competitive and Fully Cooperative Scenario}
\noindent In this paragraph, the stability of the system under either the fully competitive or fully cooperative scenarios is discussed, that is, the model with network matrix $\textbf{K}_{1}$ and $\textbf{K}_{4}$. By definition, a fixed 
point is a point in the phase space where all the time derivatives are zero, i.e., $\dot{s}_{i} = 0$ for $i = 1\ldots,n$.
The following fixed points were detected analytically from equations (\ref{4a})-
(\ref{4c}):
\begin{enumerate}
\item[(I)] $s_{i}= 0$ for $i = 1, 2, 3$. i.e. all agents with zero size.
\item[(II)]   $s_{i}= 1$ and  $s_{j}= 0$ for every $i \neq j$ i.e. one agent 
monopolizes the market.
\item[(III)] $s_{i}= b$ for $i = 1, 2, 3$, $0 < b \leq 1$. i.e. all agents own 
an equal share of the market.
\end{enumerate}

Moreover it can be easily shown that the elements of the Jacobian matrix of the 
triads are:

\begin{equation} [J]_{(i,k)}=\frac{\partial \dot{s_{i}}}{\partial s_{k}}=\left\{
                  \begin{array}{ll}
                    1 - 2s_{i} - \sum_{i \neq j} s_{j}\left(1 + 
\gamma(s_{i},s_{j})\left[1 - \frac{2}{\sigma^{2}} s_{i}(s_{i} - 
s_{j})\right]\right), & \hbox{ for $k = i$;} \\ \nonumber \\
                    -s_{i} - s_{i}\gamma(s_{i},s_{k})\left[1 - 
\frac{2}{\sigma^{2}} s_{k}(s_{i} - s_{k})\right], & \hbox{for $k \neq i$,}
                  \end{array}
                 \right  . 
\end{equation}  

from which the stability conditions are to be found at each fixed point.
\subsubsection*{Type (I) Fixed Point}
 \noindent The type (I) fixed point analysis is the case in which all agents have size zero, i.e., $s_{i}= 0$ for $i = 1, 2, 3$. The evaluated Jacobian matrix at the type (I) fixed point, is given by:
 \begin{equation} \nonumber \textbf{J} = \left[
                    \begin{array}{ccccc}
                     1 & 0& 0 \\
                      0 & 1& 0  \\
                       0&0 & 1\\
                    \end{array}
                  \right]
 \end{equation}
 \noindent whose eigenvalues are all equal to 1 (i.e.,
 $\lambda_{1} = \lambda_{2} = \lambda_{3} = 1$). Therefore, it is an unstable fixed point. At this fixed point, competition or cooperation is not applicable since all agents are at level zero. In other words, these results do not depend 
on the network matrix $K$.
  \subsubsection*{Type (II) Fixed Point}
\noindent The type (II) fixed point analysis corresponds to the case in which one agent eventually monopolizes the market and satisfies the whole demand (i.e., $s_{1} = 1$); all others have size zero (i.e., $s_{2} = s_{3} = 0$). 
Evaluating the Jacobian matrix at the type (II) fixed point, we obtain:
 \begin{equation} \nonumber \textbf{J} = \left[
                    \begin{array}{ccccc}
                     -1 & -1-\phi_{12}& -1-\phi_{13} \\
                      0 &-\phi_{12}& 0 \\
                       0& 0& -\phi_{13} \\
                    \end{array}
                  \right],
 \end{equation}
\noindent where $\phi_{ij} = k_{ij} \exp(- \sigma^{-2})$, for $i,j = 1,2,3$ and 
$i\neq j$.
The eigenvalues are obtained from
\begin{equation}
\nonumber (-1-\lambda)(-\phi_{12}-\lambda)(-\phi_{13}-\lambda) = 0
\end{equation}
which implies that $\lambda_{1} = -1$, $\lambda_{2} = -\phi_{12}$ and $\lambda_{3} = -\phi_{13}$.

In the fully competitive scenario, all $k_{ij} = 1$ for $i \neq j$, thereby resulting to an all negative eigenvalues of \textbf{J}. Indeed, this implies stability of the system at this fixed point. This is applicable in real systems: 
if an agent monopolises the competitive market, the agent will ensure  that such a domination is not lost, whence keeping the market stable.

 On the contrary, in the fully cooperative scenario, $\lambda_{2}$ and $\lambda_{3}$ are positive, since all $k_{ij} = -1, i \neq j$, thereby making the system unstable. This is also realistic, since in cooperation, the ultimate 
goal is the maximisation of all agents gain in the market. Therefore, it can be emphasized that monopoly and cooperation are not compatible terms. This accounts for  systemic  instability.

 In conclusion, the type (II) fixed point is stable in a full competitive scenario because of the possibility of monopoly, but is unstable under full cooperation.
 \subsubsection*{Type (III) Fixed Point}
 \noindent The type (III) fixed point analysis is the case in which all agents are eventually owning the same share of the market, i.e.,  $s_{i}= b$ for $i = 
1, 2, 3$, $0< b <1$.  When evaluating the Jacobian matrix at this fixed point, the constant $b$ has first to be calculated by substituting $s_{i} = b$ and $\dot{s}_{i} = 0$ into Equations (\ref{4a})-(\ref{4c}). From (\ref{4a}), it can 
be deduced that:
 \begin{eqnarray}   0 &=& b(1 - 3b)- (k_{12} + k_{13})b^{2}
 \\ \nonumber \\  &= & 1 - b(3 + (k_{12} + k_{13}).
 \end{eqnarray}
 \noindent Therefore, 
 \begin{equation}  b = \frac{1}{(3 + (k_{12} + k_{13}))}.
 \label{4h}
\end{equation}
\noindent Similarly from (\ref{4b}) and (\ref{4c}) respectively, the following is obtained:
\begin{equation}  b = \frac{1}{(3 + (k_{12} + k_{23}))}. 
\label{4i}\end{equation}
\begin{equation}  b = \frac{1}{(3 + (k_{13} + k_{23}))}.
\label{4j}\end{equation}

Therefore, from equations (\ref{4h})-(\ref{4j}), it can be deduced that in a fully competitive system (i.e. $k_{12} = k_{13}= k_{23} = 1$), the agent size is  $b = \frac{1}{5}$. This implies that the aggregate size of the three agents does 
not reach the market maximum possible capacity, which may be the negative result of the competition amongst peers.

Thus, for a full competition of agents with the same size, the Jacobian matrix with $b=\frac{1}{5}$ is:
 \begin{equation} \nonumber \textbf{J} = \frac{1}{5} \left[
                    \begin{array}{ccccc}
                     -1 & -2& -2 \\
                      -2 &-1& -2 \\
                       -2&-2 & -1\\
                    \end{array}
                  \right],
 \end{equation}
\noindent the eigenvalues are obtained from the characteristic polynomial
\begin{eqnarray}
 \nonumber(-1- \lambda)^{3} -12 (-1-\lambda) -16 &=& 0
\end{eqnarray}

 The solution to the above cubic equation is $\lambda_{1} = \lambda_{2} = 1$ and $\lambda_{3} = -5$. This shows that the system is unstable at this fixed point. 
When all the agents have an equal size in a competitive market, the system will be unstable; because the major goal of  each  agent is to $individually$ dominate the market.  According to the model, agents with ``similar" sizes are  
strongly interacting;   this leads to  a ``survival of the fittest"  scenario in  
such a competitive system, - thereby making the system unstable. 

In contrast, for a cooperative system (i.e. $k_{12} = k_{13}= k_{23}= -1$), we have $b = 1$, that is, collaboration makes all agents reach the market full capacity with agent sizes  as a function of time possibly intersecting one another.

The corresponding Jacobian matrix for the cooperative system with $b=1$ is:
 \begin{equation} \nonumber \textbf{J} =  \left[
                    \begin{array}{ccccc}
                     -1 & 0& 0 \\
                      0 &-1& 0 \\
                       0&0 & -1\\
                    \end{array}
                  \right];
 \end{equation}
\noindent the eigenvalues are obtained from the equation:
\begin{equation}
 \nonumber(-1- \lambda)^{3}= 0
\end{equation}
 \noindent which implies that  the system is stable,  since $\lambda_{1,2,3} = -1.$ When all the agents with $quasi$ equal market share cooperate, with the collective goal of maximizing their profit (size) in the market, the system will 
definitely be stable; their goal will be achieved since the strongest possible 
interaction exists amongst agents with similar sizes.

 It can be noted that when $b=\frac{1}{3}$, (i.e. $k_{12} = k_{13}= k_{23}= 0$), there exists no interaction amongst the triads agents of equal sizes and this leads to the case whereby the agents ``share the market" equally.

In conclusion to this section, a summary of the fixed point analysis of our model with the network matrix $\textbf{K}_{1}$ and $\textbf{K}_{4}$, i.e, under the fully competitive and fully cooperative scenarios is presented in Table 
\ref{table:sum}. In this table, it is  shown that systemic stability is observed under full competitive scenario only when one agent monopolizes the market, but under the fully cooperative scenario, stability occurs only when all the agents own an equal share of the market.
\begin{center}
\begin{table}[]
\centering
\begin{tabular}{|c|c|c|c|c|}
\hline
\multicolumn{4}{|c|}{Fixed Point Analysis and Stability} \\
\hline
Scenario &(I.) $s_{i}= 0$ $
\forall i$ & (II.) $s_{i}= 1$, $s_{j}=0$, $i \neq j$& (III.) $s_{i}= b$ $\forall 
i$\\
\hline
\multirow{4}{*}{Full Competition} & $\lambda_{1} = 1$  & $\lambda_{1} = -1$ & 
$\lambda_{1} = 1$ \\ & $\lambda_{2} = 1$ & $\lambda_{2} = -\phi_{12}$ & 
$\lambda_{2} =  1$ \\
& $\lambda_{3} = 1$ & $\lambda_{3} =  -\phi_{13}$ & $\lambda_{3} = -5$ \\
& Unstable & Stable & Unstable \\
\hline
\multirow{4}{*}{Full Cooperation} & $\lambda_{1} = 1$  & $\lambda_{1} = -1$ & 
$\lambda_{1} = -1$ \\ & $\lambda_{2} = 1$ & $\lambda_{2} = -\phi_{12}$ & 
$\lambda_{2} =  -1$ \\
& $\lambda_{3} = 1$ & $\lambda_{3} =  -\phi_{13}$ & $\lambda_{3} = -1$ \\  
&Unstable & Unstable & Stable \\
\hline
\end{tabular}
\label{table:sum}
\caption{Summary of analytical results under Full Competition and Full Cooperation with $\phi_{ij} = k_{ij} \exp(-\sigma^{-2})$ and $k_{ij} = \pm 1$ depending on interaction between agent $i$ and $j$.}
\end{table}
\end{center}
%\begin{tabular}{|c|c|c|c|c|}
%\hline
%\multicolumn{5}{|c|}{Fixed Point Analysis} \\
%\hline
%Scenario &Network Matrix&(I.) $s_{i}= 0$ $
%\forall i$ & (II.) $s_{i}= 1$, $s_{j}=0$, $i \neq j$& (III.) $s_{i}= b$ $\forall i$\\
%%Full Competition & $\textbf{K}_{1}$ &Unstable & Stable & Unstable \\
%%Full Cooperation &$\textbf{K}_{4}$ &Unstable & Unstable & Stable \\
%\hline
%\end{tabular}

\subsection{Mixed Interaction Scenario}
 \noindent For the mixed interaction system of triads, two possible cases  can be considered; according to the number of cooperation pairs the other cases can be easily found  isomorphic to these, i.e.,
 \begin{itemize}
 \item $G_{2}$: Agent $A_{1}$ is competing both with agent $A_{2}$ and agent 
$A_{3}$, these two being in one cooperation scheme, i.e., the model with network matrix $\textbf{K}_{2}$. This implies that in equations (4.1)-(4.3),  $k_{12} = 
1$, $k_{13} = 1$ and $k_{23} = -1$.
 \item $G_{3}$:  Agent $A_{1}$ cooperated with both agent $A_{2}$ and agent 
$A_{3}$, but these two are in competition, i.e., the model with network matrix 
$\textbf{K}_{3}$. Therefore,  $k_{12} = -1$, $k_{13} = -1$ and $k_{23} = 1$  for 
equations (\ref{4a})-(\ref{4c}).
\end{itemize}
Analytically, only one fixed point was detected for the mixed interaction of triad agents which is a case of market duopoly. The coordinates of the fixed points are given by:\\
  \noindent $s_{i}= 1$, $s_{j} =1$, $s_{k} = 0$, for some $i,j,k= 1,2,3$, 
indeed, i.e. outlining the case of duopoly in a mixed interaction market.
  
\begin{figure}
    \begin{center}
	{\includegraphics[width=0.24\textwidth, angle=-90]{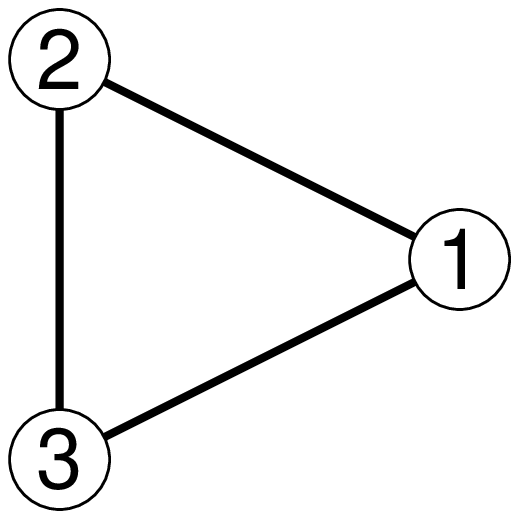}\label{subfig:graph_1}}
	{\includegraphics[width=0.24\textwidth, angle=-90]{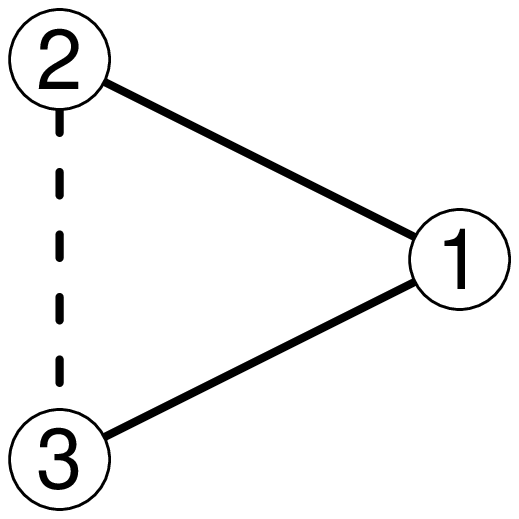}\label{subfig:graph_2}}
   {\includegraphics[width=0.24\textwidth, angle=-90]{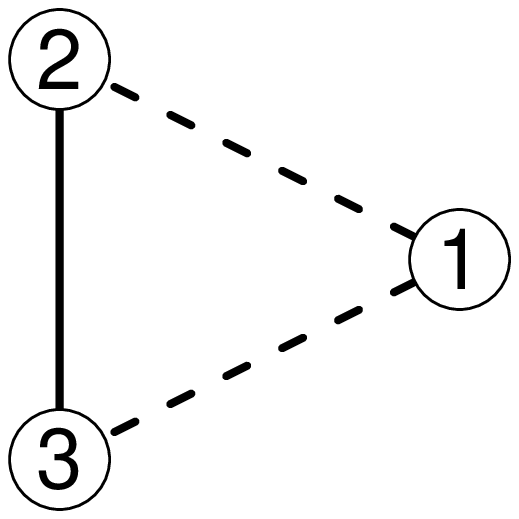}\label{subfig:graph_3}}
    {\includegraphics[width=0.24\textwidth, angle=-90]{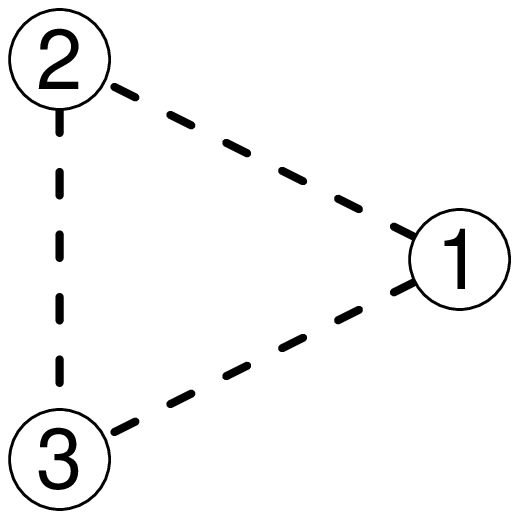}\label{subfig:graph_4}}
	\caption{Graphical illustration of the four scenarios. An edge with solid line signifies a competitive interaction while an edge with dashed line signifies a cooperative interaction. Pictures respectively refers to: (a) 
$G_{1}$, (b) $G_{2}$, (c) $G_{3}$ and (d) $G_{4}$}\label{fig:Graph}
\end{center}
\end{figure}
\subsubsection*{Fixed point Analysis}
\noindent For the fixed point analysis of the two possible mixed interaction 
scenarios, the Jacobian matrix is given by:
\begin{equation}  \textbf{J} = \left[
                    \begin{array}{ccccc}
                      -2-k_{12}& -1-k_{12}& -1-\phi_{13} \\
                      -1-k_{12} &-2-k_{12}& -1-\phi_{23} \\
                       0&0 & -1-\phi_{13} - \phi_{23}\\
                    \end{array}
                  \right],
                  \label{4k}
\end{equation}
where $\phi_{ij} = k_{ij} \exp(-\sigma^{-2})$ and $i,j =1,2,3$ for $i \neq j$. 
Considering the two possible cases of the mixed interaction scenarios, two different Jacobian matrices emerge depending on the values of each $k_{ij}$. When one agent competes  with two other agents, themselves  in cooperation, 
$k_{12} = 1$, $k_{13} = 1$ and $k_{23} = -1$ which after  substitution into (\ref{4k}) leads to the Jacobian matrix
\begin{equation} \nonumber \textbf{J} =  \left[
                    \begin{array}{ccccc}
                     -3 & -2& -1-\phi_{13} \\
                      -2 &-3& -1-\phi_{23} \\
                       0&0& -1-\phi_{13}-\phi_{23}\\
                    \end{array}
                  \right].
\end{equation}
\noindent Solving the characteristic equation
\begin{equation} \nonumber (-1-\phi_{13}-\phi_{23} - \lambda)[(-\lambda - 3)^{2} 
- 4] = 0
\end{equation}
 the eigenvalues of the Jacobian matrix is obtained to be $\lambda_{1} = -5, 
\lambda_{2}= \lambda_{3} = -1$. This signifies systemic stability under the first case of mixed interaction scenarios with the network matrix $\textbf{K}_{2}$.

When one agent cooperates with two other agents  themselves in competition with each other, $k_{12} = -1$, $k_{13} = -1$ and $k_{23} = 1$;  substitution into 
(\ref{4k}) implies that the Jacobian matrix is:
\begin{equation} \nonumber \textbf{J} =  \left[
                    \begin{array}{ccccc}
                     -1 & 0& -1-\phi_{13} \\
                      0 &-1& -1-\phi_{23} \\
                       0&0& -1-\phi_{13}-\phi_{23}\\
                    \end{array}
                  \right].
\end{equation}
\noindent The corresponding eigenvalues obtained from
\begin{equation}
 \nonumber (-1-\phi_{13}-\phi_{23} - \lambda) [(-1- \lambda)^{2}] = 0.
\end{equation}
 \noindent is $\lambda_{1,2,3} = -1$. Hence, for the fixed point analysis under second possible case of mixed interaction scenario with a network matrix $\textbf{K}_{3}$ in our model, stability is observed in the system.

In conclusion of this section, for the two types of mixed interactions amongst triad agents, systemic stability is observed; this further shows the relevance 
of examining the co-existence of competition and cooperation amongst agents in the market.

\section{SIMULATION RESULTS}
In this section, we present results  from numerical simulations emphasizing  the initial conditions and the convertgence of triad agent sets for all interesting  
and  possible  scenarios, for each network matrix  type 
$\textbf{K}=\{\textbf{K}_{1}, \textbf{K}_{2}, \textbf{K}_{3}, \textbf{K}_{4}\}$. 

For the sake of clarity, in Fig \ref{fig:Graph}, the four possible scenarios are illustrated through undirected  graphs with three vertices representing agents 
$A_{1}, A_{2}$ and $A_{3}$ and three edges which signify the type of interaction amongst the agents. An edge with solid line signifies a competitive interaction 
while an edge with dashed lines signifies a cooperative interaction. Therefore, it can be deduced that in Fig \ref{fig:Graph}, graph $\textbf{G}_{1}$ represents 
the full competitive scenario, $\textbf{G}_{2}$ represents the first case of the mixed-interactive scenario in which agent $A_{1}$ competes with $A_{2}$ and $A_{3}$, themselves in cooperation. Graph $\textbf{G}_{3}$ represents the second 
case of the mixed-interactive scenario in which agent $A_{1}$ cooperates with $A_{2}$ and $A_{3}$, themselves in competition and $\textbf{G}_{4}$ represents the full cooperative scenario.

We have tested different sets of initial conditions; see a few exemplary cases in Table \ref{table:nonlin}. We have verified the coherence of results. This suggested to us to present only cases when the initial conditions of agents 
sizes are rather different or quite similar, assuming a constant parameter that controls the size similarity, $\sigma =1$. The dynamic change in agent's size and relative behavior have been observed for each scenario. Finally, note that 
agent's sizes initial conditions were chosen within a small interval in order to allow some ``meaningful"  interaction amongst the agents; since  within our 
model, indeed, as $\mid s_{i} - s_{j}\mid \rightarrow \infty$, the interaction 
function $\gamma(s_{i},s_{j}) \rightarrow 0$.
\subsection{Fully Competitive Scenario ($G_{1}$)}
\begin{figure}
	\includegraphics[width=0.5\textwidth, angle=-90]{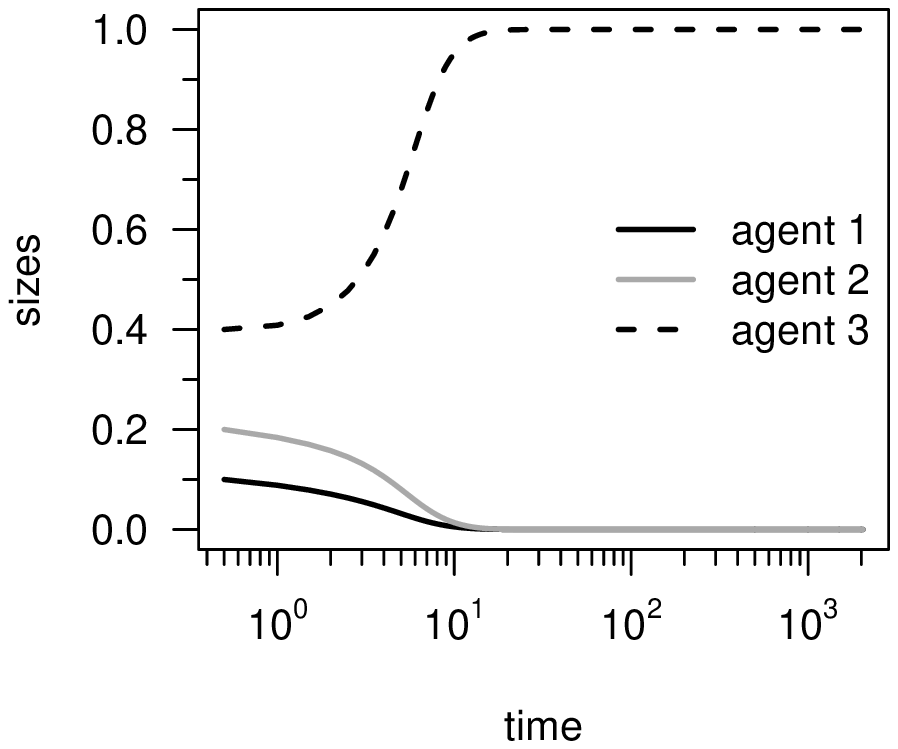}
\includegraphics[width=0.5\textwidth, angle=-90]{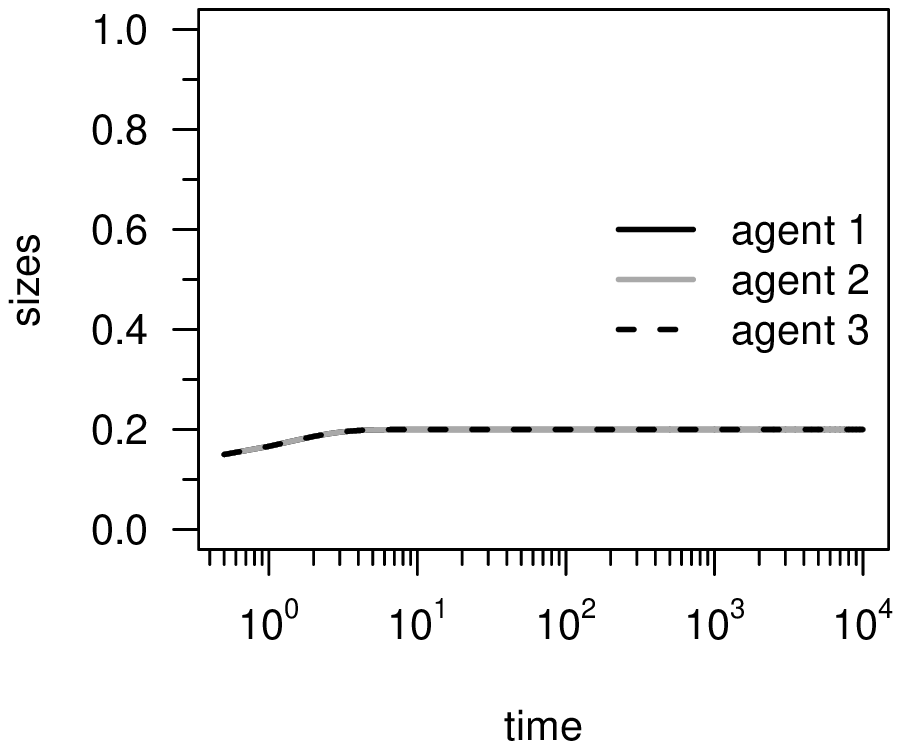}
	\caption{Fully Competitive Scenario ($G_{1}$) with  different (lhs) and similar (rhs) agent's initial sizes.}\label{fig:competitive scenario}
\end{figure}
\noindent  The fully competitive scenario with different initial conditions for the agent's size is observed in Fig \ref{fig:competitive scenario} with a 
consideration on dynamical change in the agent size; the market eventually ends in a monopoly. For all the permutations of the initial conditions, the agent that starts with the highest initial size, i.e $s_{0} = 0.4$, eventually monopolizes the market by attaining the market capacity, while the other two agents fade out of the market. This is possible because the two agents with smaller sizes compete too weakly with the agent that eventually dominates the market.

 On the contrary, when agents sizes are similar in Fig \ref{fig:competitive scenario}, the competition becomes very ``fierce" and all agents are struggling for their survival in the market. After some  time span, it is observed that the 
agents eventually have an equal share of the market;  their total market share is  however lower than the market capacity. Thus, strong competition among peers is shown to  lead to a reduction in the aggregate output due to the selfish 
interest of the individual agents. Indeed, observe that the final state ($\sum s_{i} = 0.6$). Convergence is generally slower in the competitive scenario when 
compared with the other scenarios due to the nature and effect of competition amongst the interacting agents.
\subsection{Fully Cooperative Scenario ($G_{4}$)}
\begin{figure}
	\includegraphics[width=0.5\textwidth, angle=-90]{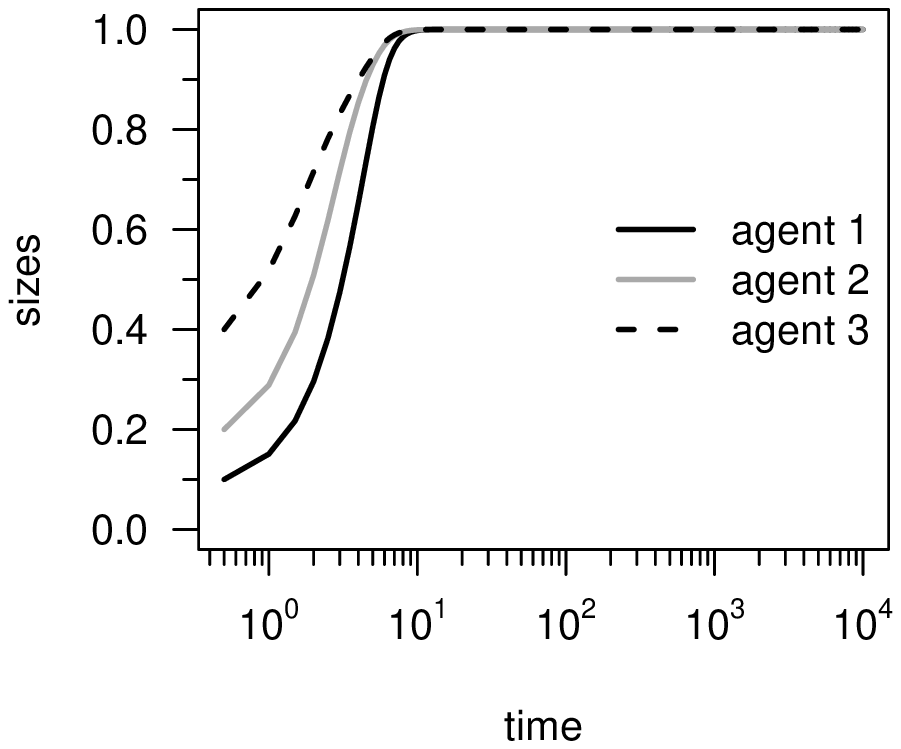}
	\includegraphics[width=0.5\textwidth, angle=-90]{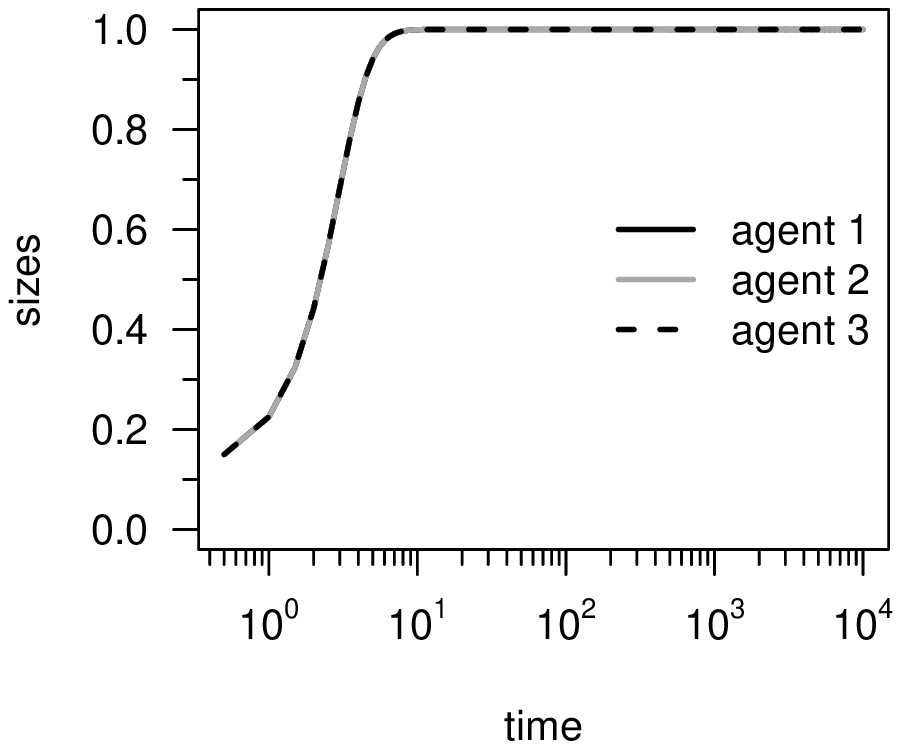}
	\caption{Fully Cooperative Scenario ($G_{4}$) with different (lhs) and similar (rhs) agent's sizes.}\label{fig:cooperative scenario}
\end{figure}
\noindent The fully cooperative scenario ($G_{4}$) is analyzed for  different initial sizes  of the triads and  illustrated in Fig \ref{fig:cooperative scenario}. The cooperation amongst the agents enables all agents to  grow in size up to this market capacity, thereby increasing the total market size, - 
which is the essence of cooperation $(\sum s_{i}> 1.0)$. A simple  analogy  can be  drawn with the case of publishing research in journals with a demand of just three papers in a journal edition (i.e. market capacity). For simplicity,  let 
it be assumed that the three authors have the same quality standard (i.e. initial condition), under full competition, each author will have one article published making up three papers. However, the ``best" situation occurs under 
full cooperation when  each author cooperates in writing the articles; they eventually have three papers each published to their credit. Hence, the final result will be three papers for the editor and three papers for each author. One 
other example pertains to car owners who may own more than one car, e.g., three cars from three different manufacturers. These simple examples show that the agent's sizes can intersect each other during full cooperation.

The simulation  leads to the same effect, even if the initial conditions are permuted amongst the triads and when the agents have  similar initial sizes as shown in Fig \ref{fig:cooperative scenario}. Also, agents tend  to "quickly 
agree", thereby converging within a short time lapse.
\subsection{The single pair cooperation ($G_{2}$).}
\begin{figure}
\includegraphics[width=0.5\textwidth, angle=-90]{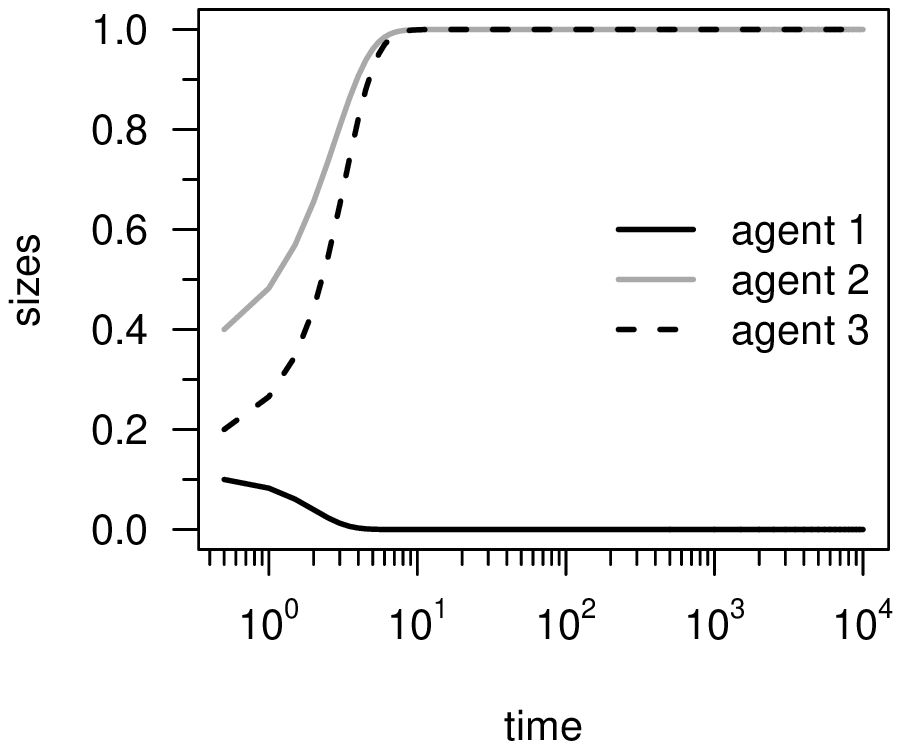}
\includegraphics[width=0.5\textwidth, angle=-90]{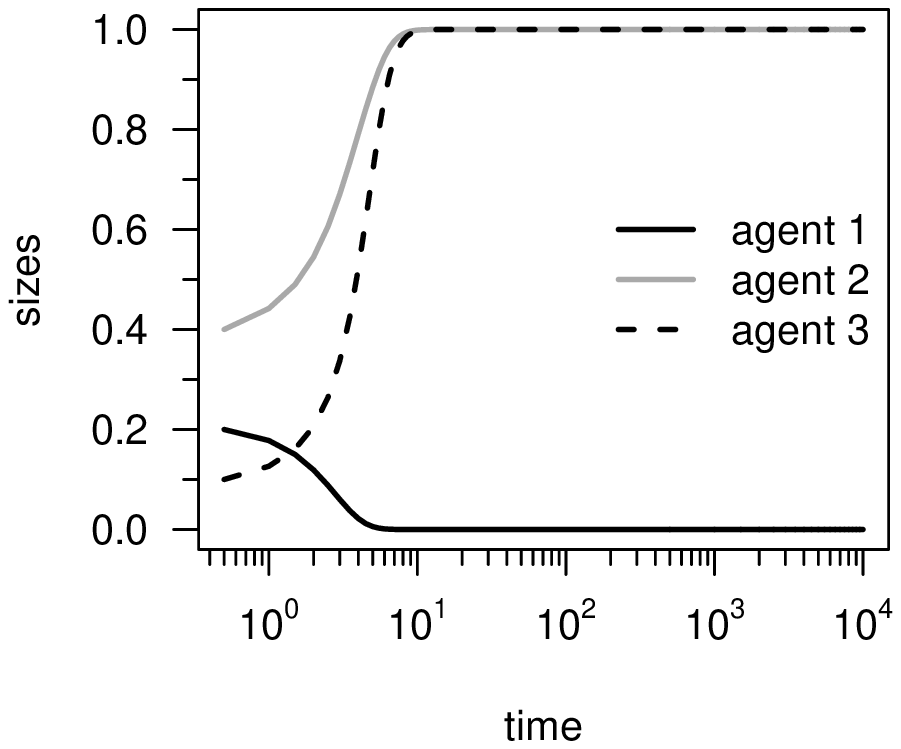}
\caption{ Mixed Interaction Scenario ($G_{2}$), $A_{1}$ competes with $A_{2}$ 
and $A_{3}$ in pair cooperation, for different initial sizes.}\label{fig:mixed A}
\end{figure}
\begin{figure}
\includegraphics[width=0.5\textwidth, angle=-90]{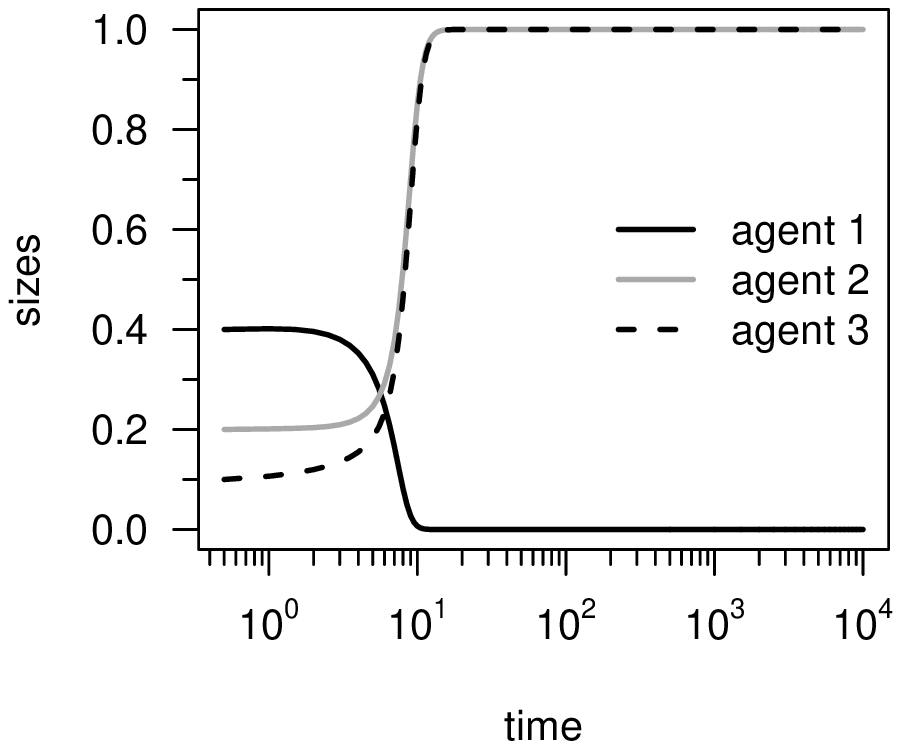}
\includegraphics[width=0.5\textwidth, angle=-90]{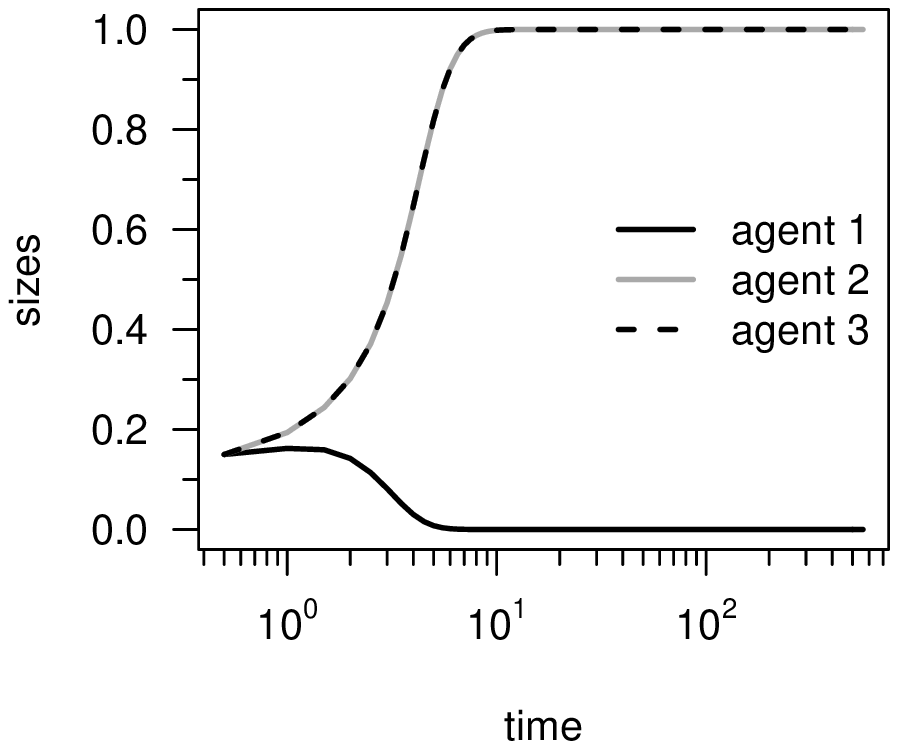}
\caption{ Mixed Interaction Scenario ($G_{2}$):  $A_{1}$ competes with $A_{2}$ and $A_{3}$ in a pair cooperation, for different (lhs) or similar (rhs) initial sizes.}\label{fig:mixed B}
\end{figure} %F5

\noindent The mixed interaction scenario ($G_{2}$), when agent  $A_{1}$ competes with agents $A_{2}$ and $A_{3}$  themselves in cooperation is shown in Fig \ref{fig:mixed A} - Fig \ref{fig:mixed B} for  different permuted initial 
conditions of agent's sizes. It can be observed that in all simulations, the sizes of agents $A_{2}$ and $A_{3}$ in cooperation grow dynamically with time up to the market capacity but agent $A_{1}$ decreases in size until it fades out of the market. The most interesting simulation is seen in Fig \ref{fig:mixed B}, with different initial conditions $s_{0} = \{0.4,0.2,0.1\}$ where agent $A_{1}$ 
initially possesses the biggest share of the market with $s_{1} = 0.4$, competes with the two other agents $A_{2}$ and $A_{3}$ that alternate a smaller 
size  $0.1$ and $0.2$. Note that the sum of the two initial shares of the cooperating agents is lower than the share of the competing agents. 
Interestingly, after  some  time, the collaboration between agents $A_{2}$ and $A_{3}$ knocks out agent $A_{1}$ from the market. An example of this scenario was experienced in  2015 in the Nigerian political history, where a political 
party that ruled the country for 16 years after democracy was restored, was defeated in a tight competition between the incumbent president and an aspirant that emerged from a strategic collaboration of three smaller political 
parties\cite{nig}. However, this is not possible if the intensity of interactions is very low amongst the agents; that is, when $A_{1}$  is  ``extremely bigger" in size when compared to agents $A_{2}$ and $A_{3}$.

When all the agents have similar initial conditions, the pattern is similar with agents $A_{2}$ and $A_{3}$ totally taking over the market by growing up to 
the market capacity as seen in Fig \ref{fig:mixed B}.
\subsection{The double pair cooperation ($G_{3}$)}
\begin{figure}
\includegraphics[width=0.5\textwidth, angle=-90]{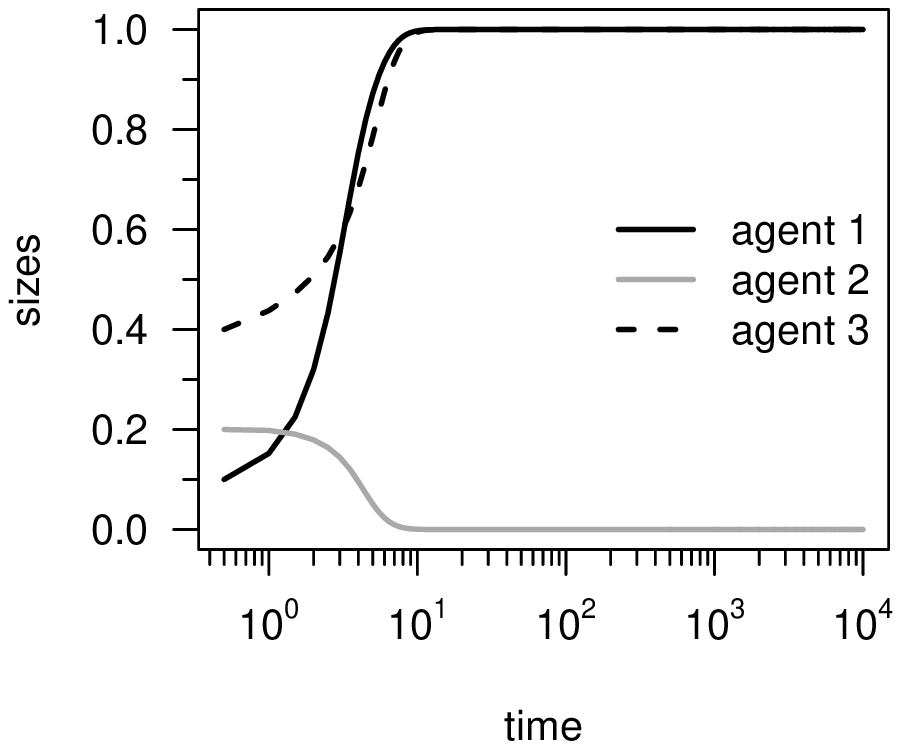}
\includegraphics[width=0.5\textwidth, angle=-90]{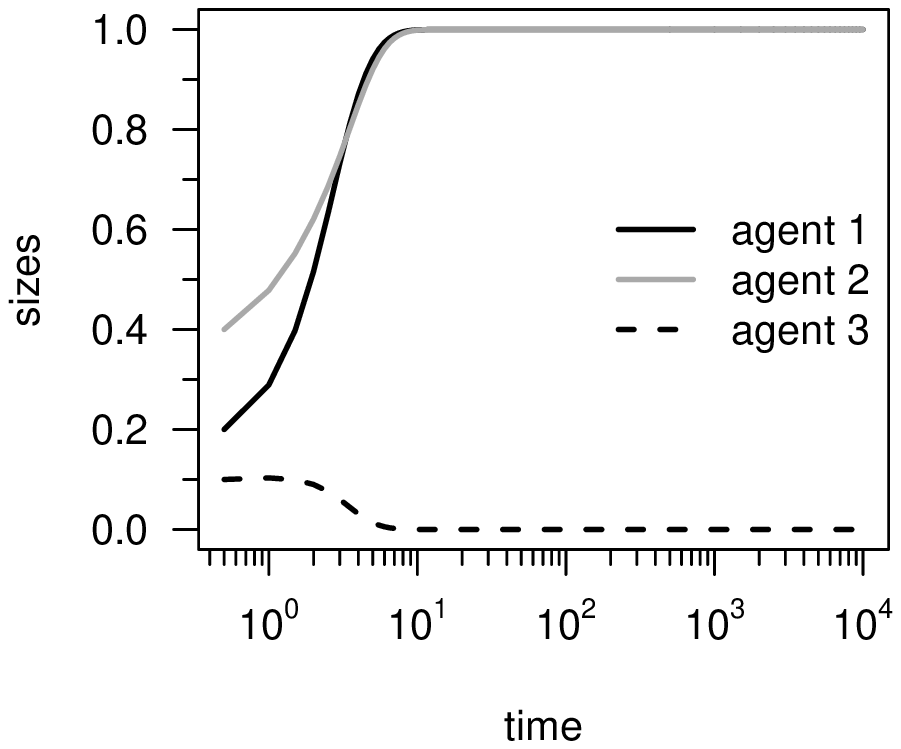}
\caption{Mixed Interaction Scenario ($G_{3}$), $A_{1}$ cooperates with $A_{2}$ 
and $A_{3}$ in competition for different  initial sizes.}\label{fig: mixed 2A}
\end{figure} %F6 
\begin{figure}
\includegraphics[width=0.5\textwidth, angle=-90]{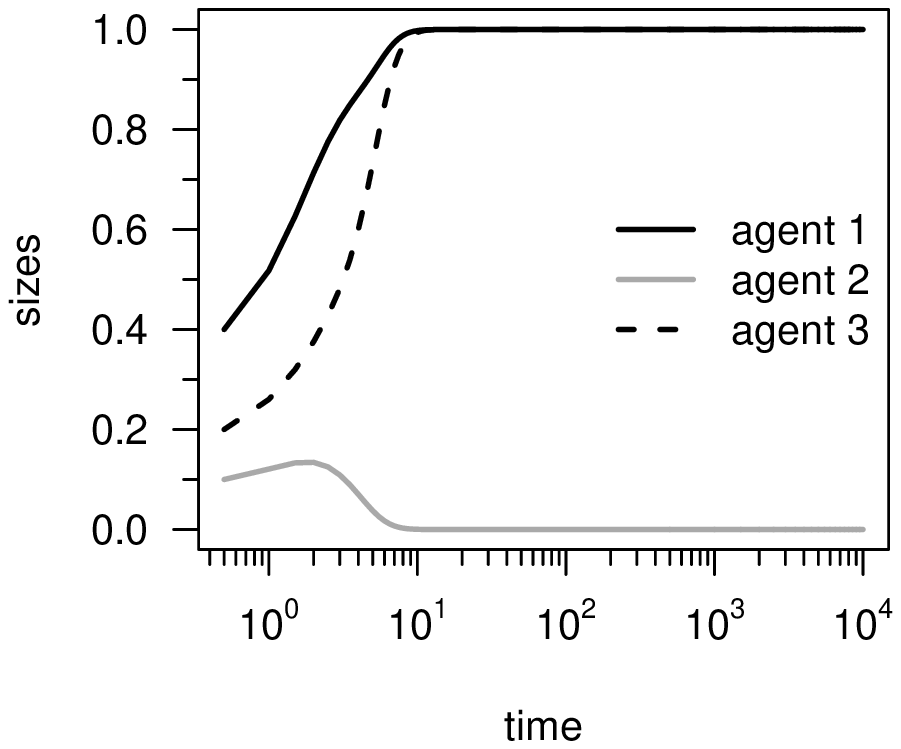}
\includegraphics[width=0.5\textwidth, angle=-90]{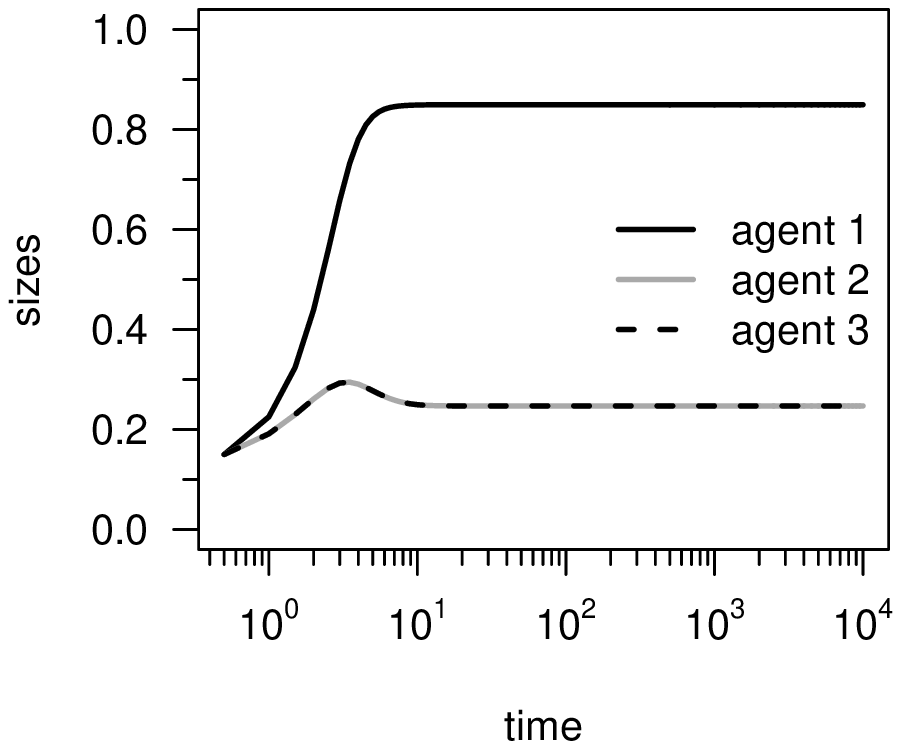}
\caption{Mixed Interaction Scenario($G_{3}$), $A_{1}$ cooperates with $A_{2}$ and $A_{3}$ in competition for different (lhs) or similar (rhs) initial 
sizes.}\label{fig: mixed 2B}
\end{figure} %F7

\noindent Finally, the second possible case of mixed interaction scenario ($G_{3}$) is shown in Fig \ref{fig: mixed 2A} - Fig \ref{fig: mixed 2B}; agent 
$A_{1}$ cooperates with agents $A_{2}$ and $A_{3}$  themselves in competition, for different permuted initial conditions of agent's sizes.
 
\begin{table}[t] \begin{centering}
	\begin{tabular}{|c|ccc|ccc|} % centered columns (4 columns)
\hline  %inserts double horizontal lines
Scenario & \multicolumn{3}{|c|}{Initial Sizes $s_i(0)$} & 
\multicolumn{3}{|c|}{Growth (+) / Decay (-)} \\ [0.5ex] % inserts table %heading
\hline
\multirow{2}{*}{Full Competition $G_1$} & 0.1 & 0.2 & 0.4 & \qquad - \qquad & \qquad - \qquad & \qquad + 
\qquad\\ & 0.1 & 0.4 & 0.2 & \qquad - \qquad &\qquad  + \qquad & \qquad  - \qquad\\
\multirow{5}{*}{\includegraphics[width=0.2\textwidth, angle=-90]{Fig_1a} }& 0.4 & 0.1 & 0.2 & \qquad + \qquad & \qquad - \qquad & \qquad - 
\qquad\\
& 0.4 & 0.2 & 0.1 & \qquad + \qquad & \qquad - \qquad & \qquad - \qquad\\ 
& 0.2 & 0.4 & 0.1 & \qquad - \qquad & \qquad + \qquad & \qquad - \qquad\\
& 0.2 & 0.1 & 0.4 & \qquad -\qquad  & \qquad - \qquad & \qquad + \qquad\\
		& 0.15 & 0.15 & 0.15  & \qquad + \qquad & \qquad + \qquad & \qquad + 
\qquad\\
		& 0.3 & 0.3 & 0.3  & \qquad - \qquad & \qquad - \qquad & \qquad - 
\qquad\\
		\hline
		\multirow{2}{*}{
			Mixed Interaction $G_2$ 		}
		& 0.1 & 0.2 & 0.4 & \qquad - \qquad & \qquad + \qquad & \qquad + 
\qquad\\
		& 0.1 & 0.4 & 0.2 & \qquad - \qquad & \qquad + \qquad & \qquad + 
\qquad\\
		\multirow{5}{*}{
			\includegraphics[width=0.2\textwidth, angle=-90]{Fig_1b} 		}
		& 0.4 & 0.1 & 0.2 & \qquad - \qquad & \qquad + \qquad & \qquad + 
\qquad\\
		& 0.4 & 0.2 & 0.1 & \qquad - \qquad & \qquad + \qquad & \qquad + 
\qquad\\
		& 0.2 & 0.4 & 0.1 & \qquad - \qquad & \qquad + \qquad & \qquad + 
\qquad\\
		& 0.2 & 0.1 & 0.4 & \qquad - \qquad & \qquad + \qquad & \qquad + 
\qquad\\
		& 0.15 & 0.15 & 0.15 & \qquad - \qquad & \qquad + \qquad & \qquad + 
\qquad\\
		& 0.3 & 0.3 & 0.3 & \qquad - \qquad & \qquad + \qquad & \qquad + 
\qquad\\
		\hline
		\multirow{2}{*}{
			Mixed Interaction $G_3$ 		}
		& 0.1 & 0.2 & 0.4 & \qquad + \qquad & \qquad - \qquad & \qquad + 
\qquad\\
		& 0.1 & 0.4 & 0.2 & \qquad + \qquad & \qquad + \qquad & \qquad - 
\qquad\\
		\multirow{5}{*}{
			\includegraphics[width=0.2\textwidth, angle=-90]{Fig_1c} 		}
		& 0.4 & 0.1 & 0.2  & \qquad + \qquad & \qquad - \qquad & \qquad + 
\qquad\\
		& 0.4 & 0.2 & 0.1  & \qquad + \qquad & \qquad + \qquad & \qquad - 
\qquad\\
		& 0.2 & 0.4 & 0.1  & \qquad + \qquad & \qquad + \qquad & \qquad - 
\qquad\\
		& 0.2 & 0.1 & 0.4  & \qquad + \qquad & \qquad - & \qquad + \qquad\\
		& 0.15 & 0.15 & 0.15 & \qquad + \qquad & \qquad + \qquad & \qquad + 
\qquad\\
		& 0.3 & 0.3 & 0.3 & \qquad + \qquad & \qquad - \qquad & \qquad - 
\qquad\\
		\hline
		\multirow{2}{*}{
			Full Cooperation $G_4$ 		}
		& 0.1 & 0.2 & 0.4  & \qquad + \qquad & \qquad + \qquad & \qquad + 
\qquad\\
		& 0.1 & 0.4 & 0.2  & \qquad  + \qquad & \qquad + \qquad & \qquad + 
\qquad\\
		\multirow{5}{*}{
			\includegraphics[width=0.2\textwidth, angle=-90]{Fig_1d}
		}
		& 0.4 & 0.1 & 0.2  & \qquad + \qquad & \qquad + \qquad & \qquad + 
\qquad\\
		& 0.4 & 0.2 & 0.1  & \qquad + \qquad & \qquad + \qquad & \qquad + 
\qquad\\
		& 0.2 & 0.4 & 0.1  & \qquad + \qquad & \qquad + \qquad & \qquad + 
\qquad\\
		& 0.2 & 0.4 & 0.1  & \qquad + \qquad & \qquad + \qquad & \qquad + 
\qquad\\
		& 0.15 & 0.15 & 0.15 & \qquad + \qquad & \qquad + \qquad & \qquad + 
\qquad\\
		& 0.3 & 0.3 & 0.3 & \qquad + \qquad & \qquad + \qquad & \qquad + 
\qquad\\
		%[1ex] % [1ex] adds vertical space
		\hline
	\end{tabular} \caption{Summary  of the effect of initial size conditions for the various scenarios  as obtained from simulations, i.e. changing the relative initial sizes of the agents}
  \end{centering}	\label{table:nonlin} % is used to refer this table in the text
\end{table}

When agent $A_{1}$ cooperates  strongly with the agent with a  higher initial condition  and
cooperates weakly with the other one. The effect of cooperation makes agent 
$A_{1}$ (irrespective of its initial condition) and any one of the two agents that cooperates strongly with agent $A_{1}$ grows up to the market capacity,
while the effect of weak cooperation and competition between agents $A_{2}$ and $A_{3}$ makes the agent with
the lowest initial condition vanishing from the market.

Surprisingly, when the three agents have the same initial size, the simulation turns out interesting also, as seen in Fig \ref{fig: mixed 2B}, - unlike the first case of mixed interaction.    When agent $A_{1}$ cooperates strongly with 
agents $A_{2}$ and $A_{3}$ which are in strong competition with each other, the effect of this strong and opposite interaction in the system results in a final state in which no agent is attaining the market capacity, but each agent 
nevertheless grows above its initial size;  all  three agents remained active in the market.

 A summary of the simulations of the triad interaction agents is presented in Table 1. The characteristic values pertain to agent's size initial conditions, final size and the time of convergence. It can be observed that convergence is
slower during competition,  due to the conflicting interests of interacting agents; in contrast, during cooperation, agents tend to ``agree", thereby converging within a shorter   time lapse. For the mixed interaction cases, when agent $A_{1}$ competes with agents $A_{2}$ and $A_{3}$ in cooperation, a 
scenario with two collaborating  agents in conflict with one, the time of convergence is seen to be faster than when agent $A_{1}$ cooperates with agents $A_{2}$ and $A_{3}$ in competition: this can be understood  due to the 
collaborative effect of mixed interactions  being higher in the first case than in the second case.

\subsection{Emphasis on the Double pair Cooperation Scenario  ($G_{3}$) with 
Non-smooth Evolutions}\label{bumps}
 
In conclusion of this simulation results section, we would like to further emphasize the two most interesting cases  $G_{3}$ and $G_{4}$, 
from a qualitative point of view in this subsection, and for a practical point of view, considering a time scale reasoning, 
in the next subsection, \ref{timescale}. 

 It can be usually noticed that the size evolution is rather smooth, - in fact remembering the (positive or negative)  logistic behavior found  
 from Verhulst equation. However, it is worth emphasizing that the double pair cooperation ($G_{3}$) can lead to a non-smooth evolution; 
 see Fig \ref{fig: mixed 2B}. Observe the figure with different initial sizes, in particular, the evolution of agent  $A_{2}$  which starts 
 with the lowest initial size. Its size first increases, but reaches a maximum; thereafter,  due to the cooperation of  agent  $A_{1}$ and agent 
  $A_{3}$, agent  $A_{2}$ is removed from the market, - even though agent  $A_{1}$  (which started with the biggest size) cooperates with agent
    $A_{2}$. {\it In fine}, $A_{1}$  ``prefers to cooperate" with  $A_{3}$, - and eliminate agent  $A_{2}$, because $A_{1}$  and $A_{3}$, 
    starting in the most advantageous positions.

 This situation does not need to be explicitly illustrated with many practical cases; it 
occurs of course in economic markets, like when two top soda companies (Coca-Cola and Pepsi-Cola, sorry for such a publicity) wish to control a country market. The same occurs in scientific competition: a well known case of war between (USA) famous laboratories occurred after the discovery of the high  $T_c$ superconducting ceramics \cite{HighTcwar}. In sport, cooperation (within theoretical competition) in order to win a race leads to temporary cooperation; see cyclism races \cite{SSJ8.91.341bicyclecoopcompet,complexityracing} or sumo  wrestling \cite{sumocompet} competitions. Competition AND cooperation between political parties in order to form a coalition have already been mentioned.
 
The other interesting case is found when all agents start with same sizes in Fig. \ref{fig: mixed 2B}.  They all start to grow, but the two competing agents with each other, agent  $A_{2}$ and agent  $A_{3}$,   loose their impetus to agent  $A_{1}$ however cooperating with both of them. In this case, agent $A_{2}$  and agent  $A_{3}$  are $not$ removed from the market,  but only reach some level yet above their initial size.  Of course, in so doing, agent  $A_{1}$ does not reach the full market capacity.

This  ``bumpy behavior" is reminiscent of the behavior found when, in Verhulst equation, the growth rate and/or the capacity are/is time dependent \cite {IJCAS30.14GompertzVerhulst}. Such a mapping into time dependent extensions of the parameters in Verhulst equation is of course outside the aim of the present paper and is left for further work.
 
\subsection{Time scale effects} \label{timescale}

Even though qualitative aspects of cooperation/competition behavior seem well described, it seems of interest to discuss whether the control parameters of the size evolution are reasonable. First recall that $\alpha$ in 
Eq.(\ref{2a}) was chosen to be equal to 1. This growth rate parameter takes values of the order of 0.04  $yr^{-1}$ for  steady populations, like in USA, e.g., in the last century. Indeed, one could estimate that the possible 
bearing time of a woman was about 25 years, and during that time she would bear one child who would be surviving. Therefore in a strict Verhulst model, the rise 
in (population) size, going roughly from 0 to 1 on the size axis, and from 1 to 10,  on the simulation scale time axis, according to all the presented figures, 
proposes a growth rate $\simeq 1/10$, - thus roughly twice as large as in reality.

It is easy to observe that the time scale due to the competition parameter is rather measured through the ratio $\alpha/k_{ij}$; see Eq.(\ref{2f}). Since we have assumed for simplicity that $k_{ij}\sim K$, the previous reasoning holds, 
leading to a convincingly reasonable value of $K\simeq 1$ for most cooperation/competition scenarios in Òreal lifeÓ. It seems reasonable to consider that the simulation scale is measured in usual time spans: years to 
hours. From the various examples that we have pointed out as applications, it is obvious that the cooperation/competition mixed cases occur over different time horizons, - whence scaling the various $ k_{ij}$  strengths.
 
\section{CONCLUSION}
\noindent In this paper, the Verhulst-Lotka-Volterra model for competition and cooperation was extended through the introduction of  some notion hereby called 
``market capacity",  in place of the individual agent's own capacity, thereby imposing a limit on agent's size growth. This improves previous studies in which 
agent's capacity are modeled individually, - studies which  resulted in agents unrealistically growing above their capacity in  a fully cooperative scenario. 

Furthermore, a network effect is introduced through   undirected and weighted graphs, which enable  a mixed-type interactive scenario, i.e., competitive and cooperative scenarios, amongst the agents.

The  present model has emphasized the basic plaquette of a network,  a triad system; it was chosen as a simple yet complex representative of any   network through which some properties of the model  can be  investigated analytically. 
In addition to this, through simulations,  dynamic changes in the agent's size and relative behavior  are observed,   for all  (4 possible)  scenarios.

We have emphasized that the initial relative sizes of agents is very relevant for the evolution of the system, leading to market sharing, or sometimes removal of agents form the ``market". Interesting scenarios occur even if the initial 
size of agents are similar. When the initial sizes are quite different, the steady is of course more quickly reached. We have emphasized that in some 
scenarios, a non smooth behavior can be found. The influence of initial conditions on the co-evolution of networks has been in this respect pointed out in \cite{PhA390.11.392RL}.

We consider that the model allows to describe the evolution of various types of agents, and is a basis for investigating more complex networks. We have pointed practical cases of interest 
throughout the text; recall co-authorship behavior in scientific publications, political 
manipulations along democratic lines, the cases of sport in which individuals from different 
teams can cooperate against rivals and reach a better status, in so doing. 
As pointed out by Porter: {\it ``The presence of multiple rivals and strong incentives influences  the 
intensity of competition among firms/agents;   yet, competition and cooperation can coexist 
because they are on different dimensions or because cooperation at some levels is part of winning 
the competition at other levels"}.

Other examples where the model can be useful may be found, beside general aspects \cite{LEMitchell}, in econophysics, 
like when there is cooperation in financial or food price speculation \cite{Chowduryfoodhike}, in auction collusions 
\cite{auctioncooperation}, or in practical economic life, e.g. when two companies decide to tie up over the introduction
 of autonomous driving taxi \cite{UberHyundai}. 

Notice that cooperation can also be corruption
\cite{sumocompet},   but can also be used to resist speculation. 
The case of false,  i.e. misleading  cooperation could also be interestingly considered, but needs further work, data, and debate.

We have provided a discussion pertaining to the possible time scales and their measure in order to give some reasonable range for the parameters of the model.
 
A final warning: it must be noticed that the interactions  are time independent and occur only amongst agents with size similarities.

Beside  including time dependence of the interaction parameters, external field effects, memory \cite{RLNatComMemory} and learning \cite{PhA368.09.1849Lipowski} 
effects could be included in further studies. Moreover, the simultaneous up-dating might be challenged for a sequential up-dating in order to find more complex behaviors. In particular, we have pointed out that  destruction of  
agents may occur; in contrast network growth can allow for creation of agents. This is relevant to observe our extension of the basic VLV modeling: the  most interesting difference between agent-based herding model and Lotka-Volterra 
model is the possibility of  investigating systems with a variable number of agents.

Finally, asymmetric interactions, e.g. removing the constraint due to the square of the 
exponential in Eq.(\ref{2b}) would be highly interesting. Can it be {\it in fine} pointed 
out that the above matrices could be asymmetric, thus sometimes, with complex eigenvalues
\cite{EPJB86.13GRMAcomplex}, whence possibly cyclic behaviors. Indeed, so called alternating 
and cut-off ways of cooperation can be envisaged \cite{KaplanRuffle}. Many developments and much work are obviously ahead. 
 %\bigskip 
\\ \\ 
\noindent{\bf Acknowledgements}
\\ \\ 
   This work is part of activities in COST ACTION TD1210  and in COST Action 
TD1306, both being gratefully acknowledged  for easing networking. In particular  
MA  has benefited of  the STSM-TD1306-33054. Comments by R. Grassi are 
appreciated.

\end{document}